\documentclass[conference, 10pt]{IEEEtran}
\IEEEoverridecommandlockouts

\usepackage{cite}
\usepackage{amsmath,amssymb,amsfonts}
\usepackage{graphicx}
\usepackage{textcomp}
\usepackage{xcolor}
\def\BibTeX{{\rm B\kern-.05em{\sc i\kern-.025em b}\kern-.08em
		T\kern-.1667em\lower.7ex\hbox{E}\kern-.125emX}}
\usepackage{subfig}
\usepackage{ upgreek }
\usepackage{graphicx}
\usepackage{graphics}
\usepackage{epstopdf}
\usepackage{amsmath}
\usepackage{multirow}
\usepackage{bm}
\usepackage{enumerate}
\usepackage{mathdots}
\usepackage{todonotes}
\usepackage{algorithmicx}
\usepackage{algpseudocode}
\usepackage{algorithm}
\usepackage{verbatim}
\usepackage{balance}
\usepackage{cite}
\usepackage[hyphens]{url}
\usepackage{acro}
\usepackage{color}

\def\Ntx{N_{\rmt}^{\rm x}}
\def\Nty{N_{\rmt}^{\rm y}}
\def\Nrx{N_{\rmr}^{\rm x}}
\def\Nry{N_{\rmr}^{\rm y}}
\def\hxtn{\hat{\bx}_{t_n}}
\definecolor{purple(x11)}{rgb}{0.63, 0.36, 0.94}
\definecolor{cadmiumgreen}{rgb}{0.0, 0.42, 0.24}

\newcommand{\deh}[1]{\hspace{#1 mm}}
\usepackage{amsmath,amssymb}

\newcommand{\bbC}{{\mathbb{C}}}

\newcommand{\bbE}{{\mathbb{E}}}

\newcommand{\bbR}{{\mathbb{R}}}

\newcommand{\ba}{{\mathbf{a}}}

\newcommand{\bi}{{\mathbf{i}}}
\newcommand{\bj}{{\mathbf{j}}}

\newcommand{\bn}{{\mathbf{n}}}

\newcommand{\bp}{{\mathbf{p}}}

\newcommand{\bs}{{\mathbf{s}}}

\newcommand{\bv}{{\mathbf{v}}}

\newcommand{\bx}{{\mathbf{x}}}
\newcommand{\by}{{\mathbf{y}}}
\newcommand{\bz}{{\mathbf{z}}}

\newcommand{\bA}{{\mathbf{A}}}

\newcommand{\bF}{{\mathbf{F}}}

\newcommand{\bH}{{\mathbf{H}}}
\newcommand{\bI}{{\mathbf{I}}}

\newcommand{\bL}{{\mathbf{L}}}

\newcommand{\bN}{{\mathbf{N}}}

\newcommand{\bS}{{\mathbf{S}}}

\newcommand{\bW}{{\mathbf{W}}}
\newcommand{\bX}{{\mathbf{X}}}
\newcommand{\bY}{{\mathbf{Y}}}
\newcommand{\bZ}{{\mathbf{Z}}}
\newcommand{\rma}{{\mathrm{a}}}

\newcommand{\rmd}{{\mathrm{d}}}

\newcommand{\rmp}{{\mathrm{p}}}

\newcommand{\rmr}{{\mathrm{r}}}
\newcommand{\rms}{{\mathrm{s}}}
\newcommand{\rmt}{{\mathrm{t}}}

\newcommand{\rmx}{{\mathrm{x}}}

\newcommand{\rmz}{{\mathrm{z}}}

\newcommand{\rmD}{{\mathrm{D}}}

\newcommand{\rmF}{{\mathrm{F}}}

\newcommand{\rmV}{{\mathrm{V}}}

\newcommand{\cI}{\mathcal{I}}

\newcommand{\cK}{\mathcal{K}}
\newcommand{\cL}{\mathcal{L}}

\newcommand{\cQ}{\mathcal{Q}}
\newcommand{\cR}{\mathcal{R}}

\newcommand{\cV}{\mathcal{V}}

\newcommand{\cX}{\mathcal{X}}

\newcommand{\cZ}{\mathcal{Z}}

\newcommand{\sfT}{\mathsf{T}}

\newcommand{\btheta}{\boldsymbol{\theta}}

\newcommand{\bphi}{\boldsymbol{\phi}}

\newcommand{\bXi}{\boldsymbol{\Xi}}

\newcommand{\bPhi}{\boldsymbol{\Phi}}
\newcommand{\bPsi}{\boldsymbol{\Psi}}

\newcommand{\transp}{{\sf{T}}}

\newcommand{\thetaaz}{\theta^{\rm az}}
\newcommand{\thetael}{\theta^{\rm el}}
\newcommand{\phiaz}{\phi^{\rm az}}
\newcommand{\phiel}{\phi^{\rm el}}

\makeatletter
\def\munderbar#1{\underline{\sbox\tw@{$#1$}\dp\tw@\z@\box\tw@}}
\makeatother

\DeclareAcronym{3GPP}{
  short=3GPP,
  long=3rd generation partnership project
}
\DeclareAcronym{5G}{
  short=5G,
  long=fifth-generation
}
\DeclareAcronym{ADC}{
  short=ADC,
  long=analog-to-digital converter
}
\DeclareAcronym{AMP}{
  short=AMP,
  long=approximate message passing
}
\DeclareAcronym{AoA}{
  short=AoA,
  long=angle-of-arrival
}
\DeclareAcronym{AoD}{
  short=AoD,
  long=angle-of-departure
}
\DeclareAcronym{APS}{
  short=APS,
  long=azimuth power spectrum
}
\DeclareAcronym{AV}{
  short=AV,
  long=autonomous vehicle
}
\DeclareAcronym{BS}{
  short=BS,
  long=base station
}
\DeclareAcronym{BSM}{
  short=BSM,
  long=basic safety message
}
\DeclareAcronym{CDF}{
	short=CDF,
	long=cumulative distribution function
}
\DeclareAcronym{CP}{
  short=CP,
  long=cyclic-prefix
}
\DeclareAcronym{CS}{
  short=CS,
  long=compressed sensing
}
\DeclareAcronym{DFT}{
  short=DFT,
  long=discrete Fourier transform
}
\DeclareAcronym{DL}{
  short=DL,
  long=downlink
}
\DeclareAcronym{DNN}{
  short=DNN,
  long=deep neural network
}
\DeclareAcronym{DoA}{
	short=DoA,
	long=direction-of-arrival
}
\DeclareAcronym{DoD}{
	short=DoD,
	long=direction-of-departure
}
\DeclareAcronym{DSRC}{
  short=DSRC,
  long=dedicated short-range communication
}
\DeclareAcronym{EKF}{
  short=EKF,
  long=extended Kalman filter
}
\DeclareAcronym{ESPRIT}{
  short=ESPRIT,
  long=estimation of signal parameters via rotational invariance techniques
}
\DeclareAcronym{FC}{
  short=FC,
  long=fully connected
}
\DeclareAcronym{FDD}{
  short=FDD,
  long=frequency division duplex
}
\DeclareAcronym{FMCW}{
  short=FMCW,
  long=frequency modulated continuous wave
}
\DeclareAcronym{FoV}{
  short=FoV,
  long=field-of-view
}
\DeclareAcronym{GNSS}{
  short=GNSS,
  long=global navigation satellite system
}
\DeclareAcronym{GPS}{
  short=GPS,
  long=global positioning system
}
\DeclareAcronym{KF}{
  short=KF,
  long=Kalman filter
}
\DeclareAcronym{LIDAR}{
  short=LIDAR,
  long=Light detection and ranging
}
\DeclareAcronym{LOS}{
  short=LOS,
  long=line-of-sight
}
\DeclareAcronym{LPF}{
  short=LPF,
  long=low pass filter
}
\DeclareAcronym{LTE}{
  short=LTE,
  long=long term evolution
}
\DeclareAcronym{LS}{
	short=LS,
	long=least square
}
\DeclareAcronym{MOMP}{
  short=MOMP,
  long=multidimensional orthogonal matching pursuit 
}
\DeclareAcronym{MIMO}{
  short=MIMO,
  long=multiple-input multiple-output
}
\DeclareAcronym{mmWave}{
  short=mmWave,
  long=millimeter-wave
}
\DeclareAcronym{MRR}{
  short=MRR,
  long=medium range radar
}
\DeclareAcronym{MSE}{
  short=MSE,
  long=mean square error
}
\DeclareAcronym{MUSIC}{
  short=MUSIC,
  long=multiple signal classification
}
\DeclareAcronym{NLOS}{
  short=NLOS,
  long=non-line-of-sight
}
\DeclareAcronym{NR}{
  short=NR,
  long=new radio
}
\DeclareAcronym{OFDM}{
  short=OFDM,
  long=orthogonal frequency-division multiplexing
}
\DeclareAcronym{ppm}{
  short=ppm,
  long=parts-per-million
}
\DeclareAcronym{RF}{
	short=RF,
	long=radio frequency
}

\DeclareAcronym{RMS}{
  short=RMS,
  long=root-mean-square
}
\DeclareAcronym{RPE}{
  short=RPE,
  long=relative precoding efficiency
}
\DeclareAcronym{RSU}{
  short=RSU,
  long=roadside unit
}

\DeclareAcronym{RX}{
  short=RX,
  long=receiver
}
\DeclareAcronym{SNR}{
  short=SNR,
  long=signal-to-noise ratio
}
\DeclareAcronym{TDoA}{
  short=TDoA,
  long=time difference of arrival
}

\DeclareAcronym{ToA}{
  short=ToA,
  long=time of arrival
}

\DeclareAcronym{TX}{
  short=TX,
  long=transmitter
}

\DeclareAcronym{UL}{
  short=UL,
  long=uplink
}

\DeclareAcronym{ULA}{
  short=ULA,
  long=uniform linear array
}
\DeclareAcronym{URA}{
	short=URA,
	long=uniform rectangular array
}

\DeclareAcronym{V2I}{
  short=V2I,
  long=vehicle-to-infrastructure
}
\DeclareAcronym{V2V}{
  short=V2V,
  long=vehicle-to-vehicle
}
\DeclareAcronym{V2X}{
  short=V2X,
  long=vehicle-to-everything
}
\DeclareAcronym{VRU}{
  short=VRU,
  long=vulnerable road user
}

\definecolor{purple(x11)}{rgb}{0.63, 0.36, 0.94}
\definecolor{cadmiumgreen}{rgb}{0.0, 0.42, 0.24}

\usepackage{pifont}

\newcommand{\be}{\begin{eqnarray}}
\newcommand{\ee}{\end{eqnarray}}
\newcommand{\degree}{^{\circ}}

\begin{document}
	\title{Sparse Recovery with Attention: A Hybrid Data/Model Driven Solution for High Accuracy Position and Channel Tracking at mmWave}
	\author{Yun Chen$^{\dag}$, Nuria Gonz\'{a}lez-Prelcic$^{\dag}$, Takayuki Shimizu$^\ddag$, Hongshen Lu$^\ddag$, and Chinmay Mahabal$^\ddag$ \thanks{This work  has been supported in part by the National Science Foundation under Grant 2147955 and by Toyota Motor North America, Inc.} \\
		$^{\dag}$ North Carolina State University, Email: \{ychen273, ngprelcic\}@ncsu.edu \\
		$^\ddag$ Toyota Motor North America, Email: \{takayuki.shimizu, hongsheng.lu, chinmay.mahabal\}@toyota.com}
	\maketitle
	
	\begin{abstract}
		In this paper, we propose first a mmWave channel tracking algorithm based on multidimensional orthogonal 
		matching pursuit algorithm (MOMP) using reduced sparsifying dictionaries, which exploits information from channel estimates in previous frames. Then, we present an algorithm to obtain the vehicle's initial location for the current frame by solving a system of geometric equations that leverage the estimated path parameters.  Next, we design an attention network that  analyzes the series of channel estimates, the vehicle's trajectory, and the initial estimate of the position associated with the current frame, to generate a refined, high accuracy position estimate. The proposed system is evaluated through numerical experiments using realistic mmWave channel series generated by ray-tracing. The experimental results show that our system provides  a 2D position tracking error below 20 cm, significantly outperforming previous work based on Bayesian filtering. 
	\end{abstract}
	
	\section{Introduction}
	Wireless communication networks are introducing sensing into the functionalities offered to their users. High accuracy localization services are relevant for several vertical industries. In particular, highly/fully automated driving applications could be facilitated if the vehicles' positions were known  by the network with an accuracy in the order of cm \cite{Nokia5GNRPosWhitePaper2021}.
	
	One way to obtain accurate location information in mmWave networks is based on exploiting the geometric relationships between the mmWave MIMO channel parameters and the position of the vehicle \cite{ShahmansooriTWC2018}. High accuracy single shot joint localization and channel estimation for initial access in vehicular systems has been addressed in recent work (see for example \cite{chen2022joint, Yun_arXiv_2023} and references therein). Once the link has been established, both the accuracy of the channel and position estimates could be further improved. The work on joint channel and position tracking is, however, scarce, both in general and for the automotive application in particular.
	
	Channel tracking methods exploiting mmWave channel sparsity and  \ac{CS} are introduced in \cite{wu2022compressive}, where an
	\ac{EKF} exploits a known channel evolution model. 
	A Kuhn-Munkres approach for channel tracking is exploited in \cite{zhu2020ray}, while \cite{chen2021millidegree} proposes to use deep learning to refine tracking results. 
	There are also studies focusing on joint channel tracking and localization \cite{koivisto2021channel, chu2022joint, kim2022pmbm}. Filters like \ac{EKF} \cite{koivisto2021channel}, particle filters \cite{chu2022joint}, and Poisson multi-Bernoulli mixture (PMBM) filters \cite{kim2022pmbm} are exploited, which can be applied independently or interactively to the channel tracking and position tracking process. 
	
	The aforementioned methods have certain limitations when applied to vehicular systems: 1) they rely on unrealistic channel evolution models that assume a constant evolving rate which does not match practical vehicular systems; 
	2) they consider the channel as containing only \ac{LOS} and first order \ac{NLOS} paths, without specifying any mechanism to identify and discard estimated second order reflections, not exploited for localization; 3) the clock offset between  the \ac{TX} and \ac{RX} is neglected; and 4) no procedures to track both angle and delay channel parameters are provided, which are required for localization when the vehicle has a single active link to a base station (BS).
	
	In this work, we focus on channel and position  tracking in a realistic urban environment. First, we propose a low complexity channel tracking method based on \ac{MOMP}\cite{palacios2022multidimensional}. Then, we design an attention network, V-ChATNet, that provides a refined, high accuracy tracking of the vehicle's position for \ac{LOS} and \ac{NLOS} settings. The inputs to V-ChATNet are the initial location estimate obtained from a geometric mapping of the tracked channel parameters and the series of previous channel and position estimates.  It identifies the channel evolution patterns, associates the channel estimates with the localization results, and provides the location corrections to keep the location error  below $0.2$ m for $95\%$ of the time.
	
	\textbf{Notations:} $[\bx]_i$ and $[\bX]_{i,j}$ denote the $i$-th entry of a vector $\bx$ and the entry at $i$-th row and $j$-th column of a matrix $\bX$ (the same rule applies for a tensor). $\bX^{\sfT}$ and $\bar{\bX}$ are the transpose and conjugate of $\bX$. $[\bX,\bY]$ and $[\bX;\bY]$ are the horizontal and vertical concatenation of $\bX$ and $\bY$. $\bX\otimes\bY$ is the Kronecker product of $\bX$ and $\bY$.
	\vspace*{-2mm}
	\section{System Model}
	We consider a mmWave vehicular communication system where the BS is equipped with a \ac{URA} of size $N_{\rmt}=\Ntx\times \Nty$, while the vehicle has  4 smaller URAs distributed on the hardtop as in \cite{Yun_arXiv_2023}, each of them of size $N_{\rmr}=\Nrx\times \Nry$ elements. A hybrid MIMO architecture is adopted to transmit $N_s$ data streams. The $q$-th time instance of the transmitted signal is denoted as $\bs[q]\in\bbC^{N_s\times 1}$, with $\bbE[\bs[q]\bs[q]^*]=\frac{1}{N_s}\bI_{N_s}$. The hybrid precoder and combiner are defined as $\bF =\bF_{\rm RF}\bF_{\rm BB}\in \bbC^{N_{\rmt}\times N_s}$ and $\bW =\bW_{\rm RF}\bW_{\rm BB}\in \bbC^{N_{\rmr}\times N_s}$, where $\bF_{\rm RF}$ and $\bF_{\rm BB}$ are the analog and digital precoders, and $\bW_{\rm RF}$ and $\bW_{\rm BB}$ are the analog and digital combiners. The $d$-th tap of the MIMO channel with $L$ paths can be formulated as
	\begin{equation}
		\bH_d=\sum\limits_{\ell=1}^{L}\alpha_\ell f_{\rmp}\left(dT_s-(t_\ell-t_{\rm off})\right)\ba_{\rmr}({\boldsymbol\theta}_\ell)\ba_{\rmt}({\boldsymbol\phi}_\ell)^*,
	\end{equation}
	where  $t_{\rm off}$ is the unknown clock offset  between the \ac{TX} and \ac{RX},  $\alpha_\ell$ and $t_\ell$ are the complex gain and the \ac{ToA} of the $l$-th path,  $T_s$ is the sampling interval, $f_{\rmp}$ is the pulse shaping function, and $\ba_{\rmr}(\btheta_\ell)$ and $\ba_{\rmt}(\bphi_\ell)$ represent the array responses of the $l$-th path evaluated at the \ac{DoA} $\btheta_\ell$, and the \ac{DoD}  $\bphi_\ell$. Note that $\ba_{\rmr}(\btheta_\ell)=\ba_{\rmr}(\thetaaz_\ell)\otimes\ba_{\rmr}(\thetael_\ell) $, and $\ba_{\rmt}(\bphi_\ell)=\ba_{\rmt}(\phiaz_\ell)\otimes\ba_{\rmt}(\phiel_\ell) $, with $\thetaaz_\ell$, and $\thetael_\ell$ the DoA in azimuth and elevation, and $\phiaz_\ell$ and $\phiel_\ell$ the DoD in azimuth and elevation. To simplify calculations, $\btheta_\ell$ and $\bphi_\ell$ are defined as unitary direction vectors, i.e., ${\btheta}_\ell=[\cos\thetael_\ell\cos\thetaaz_\ell, \cos\thetael_\ell\sin\thetaaz_\ell, \sin\thetael]^{\transp}$; a similar definition applies to $\bphi_\ell$.
	Assuming the channel has $N_{\rmd}$ taps, the $q$-th instance of the received  signal is
	\begin{equation}\label{receive_sig}
		\by[q]=\bW^*\sum\limits_{d=0}^{N_{\rmd}-1}\sqrt{P_{\rmt}}\bH_d\bF\bs[q-d]+\bW^*\bn[q],
	\end{equation}
	where $P_{\rmt}$ is the transmitted power, and $\bn[q]\sim \mathcal{N}({\bf 0},\sigma_{\bn}^2\bI_{N_s})$ is modeled as additive white Gaussian noise (AWGN). We define the whitened received signal as $\breve{\by}[q]=\bL^{-1}\by[q]$, where $\bL$ is computed via Cholesky decomposition of $\bW^*\bW=\bL\bL^*$. This way, the resulting noise term in the whitened received signal can be modeled as AWGN, i.e. $\mathbb{E}[\breve{\bn}[q]\breve{\bn}[q]^*]=\sigma_{\bn}^2\mathbb{E}[\bL^{-1}\bW^*\bW(\bL^{-1})^{*}]= \sigma_{\bn}^2\bI$. Accordingly, the whitened received signal can be written as
	\begin{equation}\label{receive_mat}
		\breve{\bY} =\breve{\bW}^* [\bH_0,...,\bH_{N_{\rmd}-1}]\left((\bI_{N_{\rmd}}\otimes\bF )\bS \right)+\breve{\bN},
	\end{equation}
	where $[\breve{\bY}]_{:,q}=\breve{\by}[q]$, $[\breve{\bN}]_{:,q}=\breve{\bn}[q]$, and $[\bS]_{:, q}=\left[\bs[q]; \bs[q-1];...;\bs[q-(N_{\rmd}-1)]\right]$. 
	
	\section{Position and channel tracking system}\label{sec:pos-track}
	\begin{figure}[t!]
		\centering
		\includegraphics[width=.7\columnwidth]{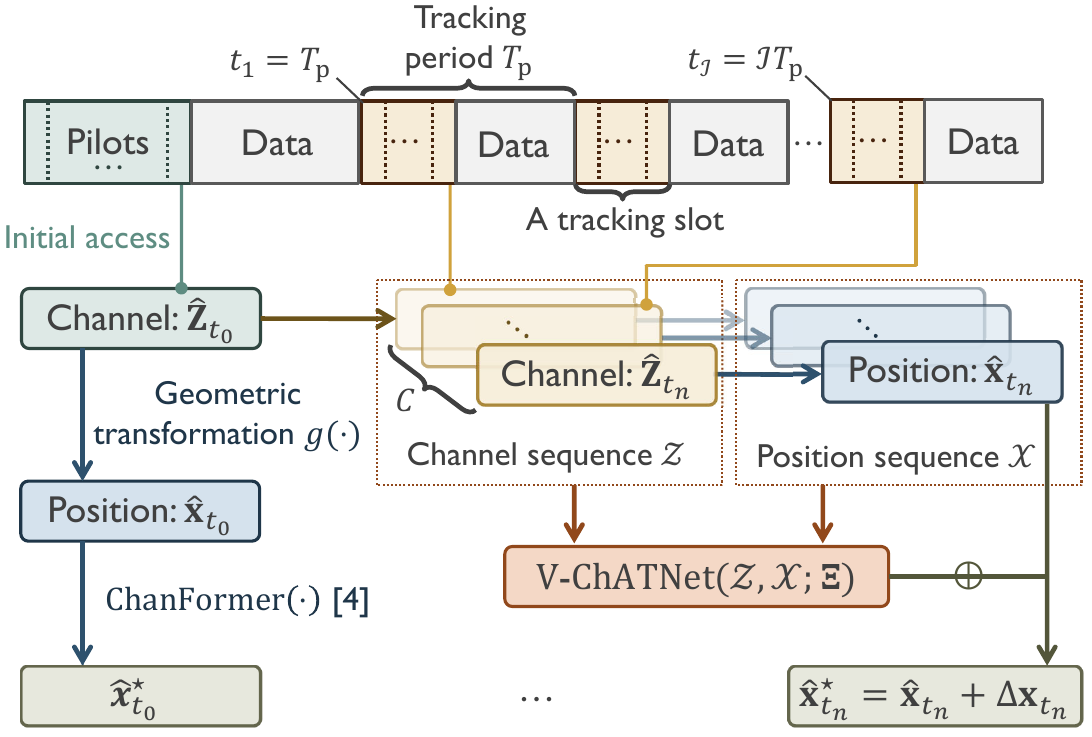}
		\caption{Diagram of the position and channel tracking system.}
		\vspace{-1.2em}
		\label{Sys_model}
	\end{figure}
	As discussed in Section I, previous work on high accuracy localization requires either delay and angular information from a single BS, or communication with several BSs to obtain an estimate of the position. In our system model we consider a communication link between a vehicle and a single mmWave BS, so delay and angular parameters of the channel need to be tracked. The design we propose in this Section tackles the problem of tracking the channel and the vehicle's position, while the channel estimation for initial access and initial localization could be realized by other methods in previous work, such as \cite{chen2022joint, Yun_arXiv_2023}.
	The block diagram of our proposed channel and position tracking system  is shown in Fig.\ref{Sys_model}. The channel tracking period is set to $T_{\rmp}$, within which the channel is considered invariant. Every  $T_{\rmp}$, the channel is  tracked using the procedure described in Section~\ref{sec:chan-track}. For every $T_n = nT_p$, the channel is fully   re-estimated using the procedure described in \cite{chen2022joint, Yun_arXiv_2023} to consider the case where significant changes occur over time  (for example, a cluster that appears or disappears).  The estimated channel is denoted as $\hat{\bZ}\in \bbR^{N_{\rm est}\times 6}$, where $N_{\rm est}$ is the number of estimated paths, and the $\ell$-th row contains the estimated parameters for the $\ell$-th path, i.e.  $\hat{\bz}_\ell=[\hat{\alpha}_\ell, \hat{\tau}_\ell, \hat{\theta}_\ell^{\rm az}, \hat{\theta}_\ell^{\rm el}, \hat{\phi}_\ell^{\rm az}, \hat{\phi}_\ell^{\rm el}]$.  We exploit the estimated \ac{TDoA} $\hat{\tau}_\ell=\hat{t}_\ell-t_{\rm off} - (\hat{t}^{\min}-t_{\rm off})=\hat{t}_\ell-\hat{t}^{\min}$ for localization instead of the \ac{ToA}, where $\hat{t}^{\min}=\min\{\hat{t}_\ell|\ell=1,...,N_{\rm est}\}$. The estimated paths have to satisfy the requirements for localization, i.e., the number of first order reflections has to be $\geq 1$ in \ac{LOS} channels or  $\geq 3$ in the \ac{NLOS} case \cite{chen2022joint}. Otherwise, the vehicle cannot be located.  We use \textit{PathNet} \cite{Yun_arXiv_2023} to determine the path orders and identify the LoS and first order reflections exploited for localization. Then, the initial 3D location estimate at the $n$-th time slot $t_n=nT_{\rmp}$ is defined as $\hat{\bx}'_{t_n}=g\left(\hat{\bZ}_{t_n}\right)$, where $g(\cdot)$ is the solution to the geometric system of equations defined in \cite{Yun_arXiv_2023}. Since we are interested in the 2D position of the cars driving on the road, an attention network, V-ChATNet, designed in Section \ref{sec:pos-track}, is then applied to refine the 2D vehicle location estimate $\hxtn$, where $\hxtn=\left[g\left(\hat{\bZ}_{t_n}\right)\right]_{:2}$ if the vehicle can be located accordingly to the aforementioned criteria. Otherwise, $\hxtn=\hat{\bx}_{t_{n-1}}+T_\rmp\hat{\bv}_{t_{n-1}}$, where $\hat{\bv}_{t_{n-1}}$ is the rough speed read from the speedometer at $t_{n-1}$. A sequence of historical channel estimates $\cZ$ and 2D location estimates $\cX$ are the input to the network, while the output is the correction of the current location estimate $\Delta\bx_{t_n}=[\Delta x, \Delta y]$. The final location estimate is computed as $\hat{\bx}^{\star}_{t_n}=\hxtn+\Delta\bx_{t_n}$. In the reminder of this Section, we describe the details of the channel tracking algorithm and the attention network for position refinement.

	\begin{figure*}[t!]
		\centering
		\includegraphics[width=0.7\textwidth]{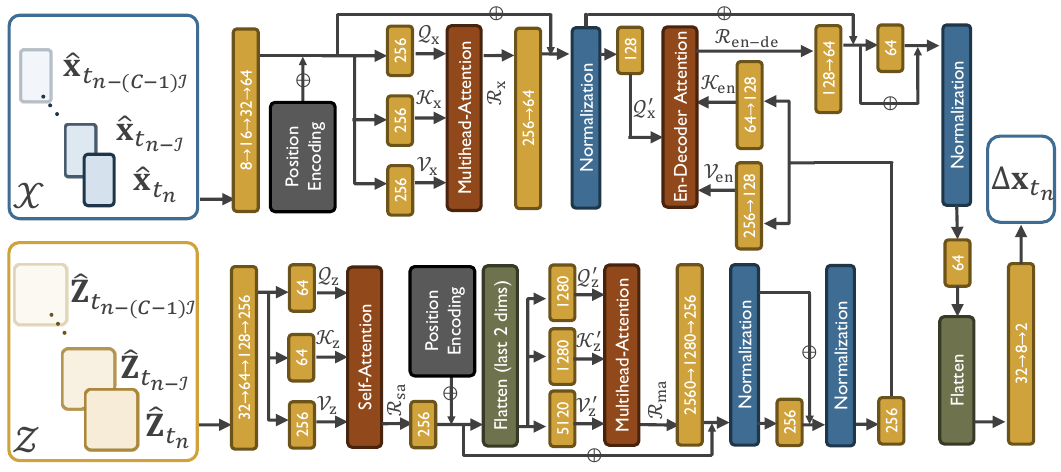}
		\caption{Architecture of V-ChATNet which takes historical channel estimates and localization results from geometric transformations as the input, and outputs the correction for the current location estimate.}
		\vspace{-1.3em}
		\label{V-ChATNet_fig}
	\end{figure*}
	\vspace*{-2mm}
	\subsection{MOMP-based Channel Tracking}\label{sec:chan-track}
	Solutions for mmWave channel tracking proposed in previous work do not consider delay tracking, 
	which disables the possibility of using these methods in a position tracking scenario where there is communication with a single BS. Second, they exploit a theoretical evolution model for the channel parameters that can be hardly met by a realistic vehicular channel, which leads to inaccurate estimations of the channel parameters. To overcome these limitations, we propose a channel tracking method that incorporates delay tracking without exploiting any rigid parameter evolution model. Our only assumption will be that the parameters will change smoothly, without considering any particular mathematical form. We will exploit this idea and the sparsity of the mmWave channel to develop a tracking procedure that relies on the recently defined \ac{MOMP} algorithm \cite{palacios2022low, palacios2022multidimensional}. This algorithm solves a sparse recovery problem for channel estimation, without relying on sparsifying dictionaries based on Kronecker products, so that it can operate with large and planar arrays without incurring in prohibitive complexity and memory requirements, as discussed in \cite{palacios2022low, palacios2022multidimensional, Yun_arXiv_2023}. The \ac{MOMP} algorithm 
	solves the optimization problem
	\begin{equation}\label{eq:MOMP}
		\min_{\bf X}\left(\left\|{\bf Y}-\sum_{{\bi}\in\mathfrak{I}}\sum_{{\bf j}\in\mathfrak{J}}[{\bf \Phi}]_{:, \bi}\left(\prod_{k = 1}^{N_{\rm D}}[{\bf \Psi}_{k}]_{i_k, j_k}\right)[{\bf X}]_{{\bf j}, :}\right\|^2_{\rmF}\right),
	\end{equation}
	where $\bY=[{\rm vec}(\breve{\bY}_1);...;{\rm vec}(\breve{\bY}_M)]$ collects $M$ measurements using different pairs of training precoders and combiners, $\bF_m$ and $\bW_m$, for every training frame $m$, with $m=1,\ldots,M$; ${\bf \Phi}\in\mathbb{C}^{MQN_s\times\otimes_{k=1}^{N_{\rm D}}N_k^{\rm s}}$, where $[\bPhi]_{(m-1)QN_s+q, \bi}=[(\bS_m^{\sfT}(\bI_{N_\rmd}\otimes \bF_m^{\sfT}))\otimes\breve{\bW}_m^*]_{q,I}$   is the measurement tensor, where $I=(i_{N_\rmD}\deh{-1}-\deh{-1}1)\prod\limits_{k=1}^{N_\rmD-1}N_k^\rms+\deh{-1}\sum\limits_{k=1}^{N_\rmD-2}\left((i_k\deh{-1}-\deh{-1}1)\prod\limits_{k'=k+1}^{N_\rmD-1}N_{k'}^\rms\right)+i_{N_\rmD-\deh{-1}1}$, $N_{\rmD}$ is the number of independent dimensions, and $N_k^\rms$ is the length of the response vector along $k$-th dimension; $\bPsi_k\in\bbC^{N_k^{\rms}\times N_k^{\rma}}$ is the dictionary along dimension $k$, where $N_k^{\rma}$ is the size of $k$-th dictionary depending on the resolutions; $\mathfrak{I}=\{\bi=[i_1,...,i_{N_{\rmD}}]\ |\ i_k\leq N^{\rms}_k\}$ and $\mathfrak{J}=\{\bj=[j_1,...,j_{N_{\rmD}}]\ |\ j_k\leq N^{\rma}_k\}$ are the sets for indices; and ${\bf X}\in\mathbb{C}^{\otimes_{k=1}^{N_{\rm D}}N_k^{\rm a}\times 1}$ is the sparse tensor whose supports indicate the entries of the dictionaries to determine the channel parameters. MOMP-based channel estimation for initial access is introduced in \cite{palacios2022low, palacios2022multidimensional, Yun_arXiv_2023}, where the angular dictionaries for estimating the \ac{DoA}/\ac{DoD} span the whole range of $[0, \pi]$, with a resolution of $\frac{\pi}{N_k^{\rma}}$, and the delay ranges from $0$ s to $N_{\rmd}T_s$, with a resolution of $\frac{N_{\rmd T_s}}{N_3^{\rma}}$. For the tracking case,  the dictionary for channel estimation at time $t_n$ can be constructed, however, with the information from $t_{n-1}$ to reduce the dictionary dimensions, reduce complexity  and facilitate the estimation process. Let the estimated channel at time $t_{n-1}$ be $\hat{\bZ}_{t'_n}$, where the $\ell$-th estimated path is  $\hat{\bz}_{t'_n,\ell}=[\hat{\alpha}_{t'_n,\ell}, \hat{\tau}_{t'_n,\ell}, \hat{\theta}_{t'_n,\ell}^{\rm az}, \hat{\theta}_{t'_n,\ell}^{\rm el}, \hat{\phi}_{t'_n,\ell}^{\rm az}, \hat{\phi}_{t'_n,\ell}^{\rm el}]$. The reduced angular dictionaries at $t_n$ are determined as:
	\begin{equation}
		\begin{cases}
			\bPsi_{1,t_n}=[\bar{\bA}(\hat{\phi}_{t'_n,1}^{\rm az})\cup...\cup\bar{\bA}(\hat{\phi}_{t'_n,N_{\rm est}}^{\rm az})] \\ 
			\bPsi_{2,t_n}=[\bar{\bA}(\hat{\phi}_{t'_n,1}^{\rm el})\cup...\cup\bar{\bA}(\hat{\phi}_{t'_n,N_{\rm est}}^{\rm el})]\\ 
			\bPsi_{3,t_n}=[\bA(\hat{\theta}_{t'_n,1}^{\rm az})\cup...\cup\bA(\hat{\theta}_{t'_n,N_{\rm est}}^{\rm az})] \\ 
			\bPsi_{4,t_n}=[\bA(\hat{\theta}_{t'_n,1}^{\rm el})\cup...\cup\bA(\hat{\theta}_{t'_n,N_{\rm est}}^{\rm el})]\\ 
		\end{cases},
	\end{equation}
	\vspace*{-0.2mm}
	where $\bA(\varphi)\hspace*{-1mm}=\hspace*{-1mm}\left\{\ba(\varphi-\omega)\hspace*{-1mm},\hspace*{-1mm}\ba\left(\varphi-\omega+\Delta\omega\right),..., \ba(\varphi+\omega) \right\}$ covers an angular sector of width $2\omega$ with a resolution of $\Delta\omega$, and $\bar{\bA}(\varphi)$ stands for the matrix that contains the conjugate of the entries in $\bA(\varphi)$. Usually, $\omega$ is selected based on the beam width, assuming that the optimal beam pair remains the same throughout the tracking period. Furthermore, the reduced delay dictionary is determined as	
		\begin{equation}
			\bPsi_{5,t_n}\deh{-1}=\deh{-1}\left[\bp_{\rmd}\deh{-1}\left(\hat{t}^{\min}_{t'_n}\right)\deh{-1}, \bp_{\rmd}\deh{-1}\left(\hat{t}^{\min}_{t'_n}\deh{-1}+\deh{-1}\Delta\tau\right)\deh{-1}, ..., \bp_{\rmd}\deh{-1}\left(\hat{t}^{\min}_{t'_n}\deh{-1}+\deh{-1}\hat{\tau}^{\max}_{t'_n}\deh{-1}+\deh{-1}\varepsilon\right)\right]\deh{-1},
		\end{equation}
	where $\Delta\tau$ is the delay dictionary resolution, $\hat{t}^{\min}_{t'_n}=\min\left\{\hat{t}_{t'_n,\ell}\ |\ \ell=1,...,N_{\rm est}\right\}$, $\hat{\tau}^{\max}_{t'_n}=\max\left\{\hat{\tau}_{t'_n,l}\ |\ l=1,...,N_{\rm est}\right\}$, $\varepsilon$ is the allowed delay extending range, and $\bp_{\rmd}(\cdot)\in\bbC^{{N_\rmd}\times 1}$ is the delay response where $[\bp_{\rmd}(t)]_{n}=f_{\rmp}\left((n-1)T_s-t\right)$. By using the reduced dictionaries, the algorithm focuses on the areas where paths are most likely to exist, thereby solving the sparse recovery problem in \eqref{eq:MOMP} with reduced complexity. In addition, the resolutions could be configured to be smaller than those used for the full dictionaries to further improve the accuracy.
	
	\subsection{V-ChATNet for Vehicle Location Tracking}
	Attention schemes have been broadly studied in prior work to address context-aware problems \cite{chaudhari2021attentive}. Our proposed V-ChATNet network enables the analysis of past channel estimates, the identification of channel evolution patterns, and the linking of the historical trajectory with the channel features for the correction of the current location estimate. The architecture of V-ChATNet is depicted in Fig. \ref{V-ChATNet_fig}. 
	
	At the encoder, the network takes in a sequence of channel estimates $\cZ=\left[\hat{\bZ}_{t_{n-(C-1)\cI}},\hat{\bZ}_{t_{n-(C-2)\cI}},...,\hat{\bZ}_{t_{n-\cI}}, \hat{\bZ}_{t_{n}}\right]\in \bbR^{C\times N_{\rm est}\times 6}$, where $C$ is the length of the input sequence, and $\cI$ is the sample interval, which should be appropriately selected to adequately reveal the channel temporal evolution features. $\cZ$ is first processed to obtain  three abstract representations as \textit{Value} $\cV_{\rmz}$, \textit{Key} $\cK_{\rmz}$, and \textit{Query} $\cQ_{\rmz}$, for self-attention of the paths of each channel. Mathematically,
	\begin{equation}\label{AttEqu}
		\cR_{\rm sa}={\rm Attention}(\cQ_{\rmz},\cK_{\rmz},\cV_{\rmz})={\rm softmax}(\frac{\cQ_{\rmz}\cK_{\rmz}^{\sfT}}{\sqrt{{\rm dim}(\cK_{\rmz})}})\cV_{\rmz},
	\end{equation}
	where the calculations are for the last dimension of the abstractions. $\cR_{\rm sa}$, which has the same shape as $\cV_{\rmz}$, is the representation where each path incorporates effects from other paths in each channel, i.e., more accurately estimated paths correspond to higher weights. Afterwards, position encoding is applied to record the chronological order of the channels. The following multi-head attention stage extracts the temporal evolution features of the channels, with greater attention given to the channels that are better estimated in the sequence. The resulting representation $\cR_{\rm ma}$ goes through the feed forward and normalization modules and transforms to $\cV_{\rm en}$ and $\cK_{\rm en}$ as the output from the encoder.
	
	The decoder takes the historical vehicle location estimates $\cX=\left[\hat{\bx}_{t_{n-(C-1)\cI}},\hat{\bx}_{t_{n-(C-2)\cI}},...,\hat{\bx}_{t_{n-\cI}}, \hat{\bx}_{t_{n}}\right]$ as the input. After feature expansions via fully connected layers, $\cX$ results in $\cV_{\rmx}$, $\cK_{\rmx}$, and $\cQ_{\rmx}$ to perform the multihead-attention to extract both the localization and the vehicle moving patterns over the given time period. The resulting representation of the location sequence $\cR_{\rmx}$ is treated as a query $\cQ'_{\rmx}$ to work with $\cV_{\rm en}$ and $\cK_{\rm en}$ for the encoder-decoder multi-head attention process. This helps establish the connections among the channel evolution, vehicle's trajectory, and the system errors introduced by the channel estimation and localization methods. The attention output $\cR_{\rm en-de}$ passes through the feed forward and normalization modules to provide the correction for the initial location estimate $\Delta\bx_{t_n}$, so that the refined location becomes $\hat{\bx}^{\star}_{t_n}=\hat{\bx}_{t_n}+\Delta\bx_{t_n}$.
	
	In summary, V-ChATNet performs a regression task, i.e., $\Delta\bx_{t_n}=\rmV\text{-}{\rm ChaTNet}(\cZ, \cX; \bXi)$, where $\bXi$ represents the network parameters to be trained. The loss function for training the network is the \ac{MSE} loss defined as $\cL(\bXi)=\|\hat{\bx}^{\star}_{t_n}-\bx_{t_n}\|^2$.

	\section{Simulation Results}
	We consider an urban canyon environment where cars and trucks are distributed across $4$ lanes and moving at the speed limits assigned to each lane: $60$, $50$, $25$, and $15$ km/h. We pick an active vehicle driving at $60$ km/h on the first lane for the tracking experiment. A $16\times 16$ \ac{URA} and $4$ $12\times 12$ URAs are deployed at the BS and the vehicle, respectively. The channel tracking period is set to $T_{\rmp}=0.5$ ms. In every tracking period, the BS transmits a data stream  drawn from a row of a Hadmard matrix of size $Q=64$, with a transmitted power of $P_{\rmt}=40$ dBm. A raised-cosine filter with a roll-off factor of $0.4$ is used as the pulse shaping function. The system operates at the carrier frequency of $f_c=73$ GHz, with a bandwidth of $B_c=1$ GHz. The dataset containing the channel series is generated by ray-tracing simulations using \textit{Wireless Insite}, which provides realistic channels with higher order reflections, much weaker \ac{NLOS} paths compared to the \ac{LOS}, etc. 
	We take $8$ scenes where the vehicles' initial positions are different, so that the dataset contains $8$ trajectories, with a split of 3:1 to form the training and testing sets for V-ChATNet. The trajectories have a length of $50$ m in average, i.e., each set consists of roughly $6000$ data samples.
	
	\begin{figure}[t!]
		\centering
		\includegraphics[width=.9\columnwidth]{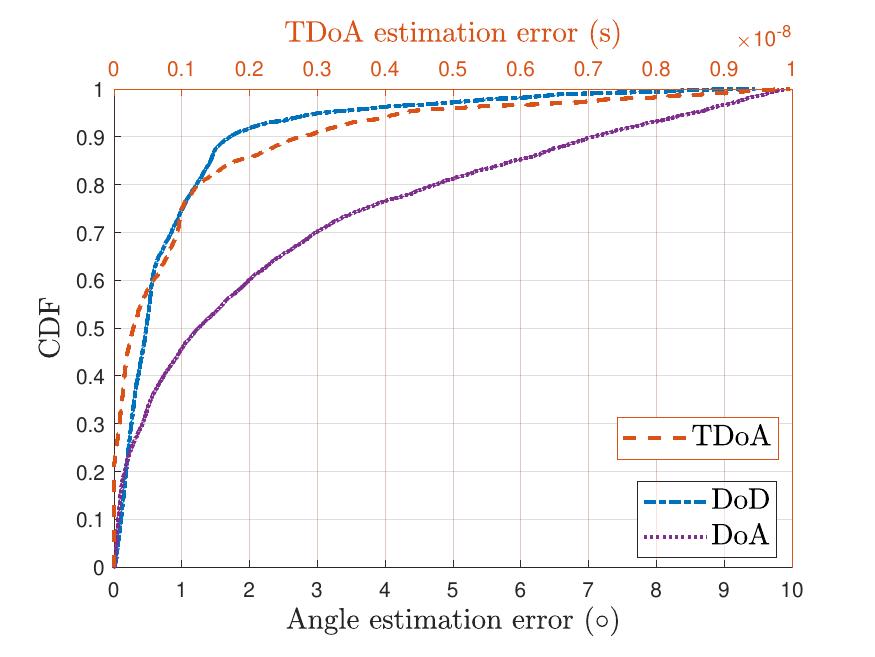}
		\caption{Performance of channel tracking using reduced dictionaries with the tracking period of $T_{\rmp}=0.5$ ms.}
		\vspace{-1em}
		\label{Perfrom_ChanTrack}
	\end{figure}
	\begin{figure}[t!]
		\centering
		\subfloat[]{%
			\label{example_traj}
			\includegraphics[width=.9\columnwidth]{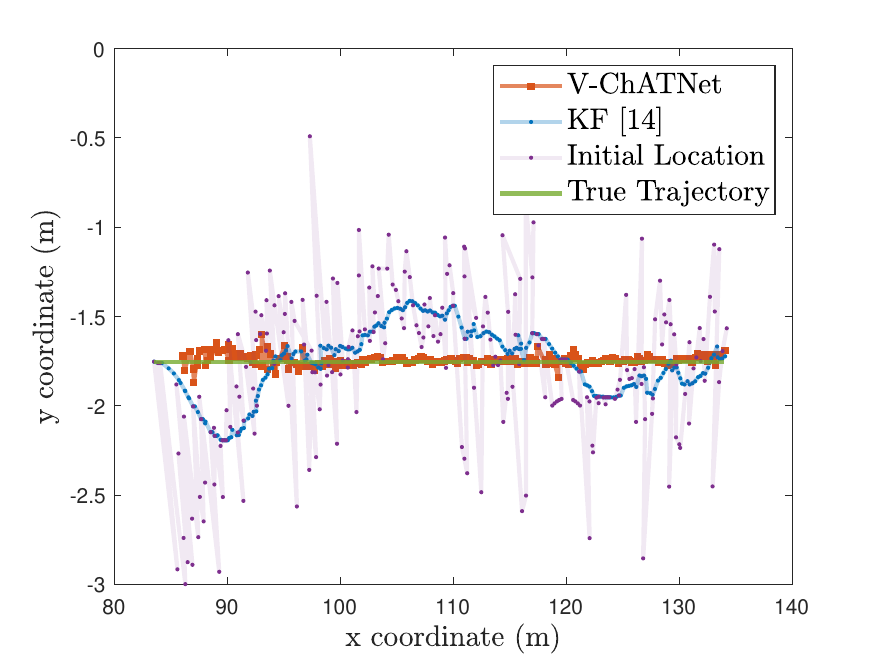}}\vspace{-0.22em}
		\subfloat[]{%
			\label{CDF_loc}
			\includegraphics[width=.9\columnwidth]{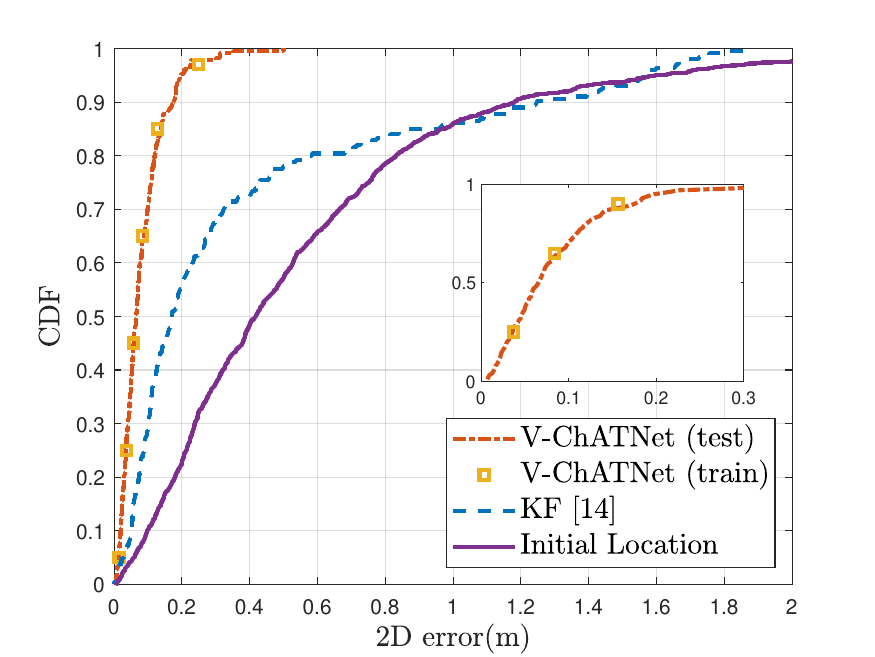}}
		\caption{Vehicle position tracking performance based on the testing set. (a) An example of tracking results from sky view; (b) Vehicle position tracking accuracy using V-ChATNet, comparing with that using KF \cite{khalkhali2020vehicle} and initial localization based on channel geometric transformations. }
		\vspace{-1.5em}
		\label{NetPerform}
	\end{figure}
	We set $\omega=15\degree$ and $\Delta\omega=0.175\degree$ for the definition of the  reduced angular dictionaries, and $\varepsilon=0.2$ ns with a resolution  of $\Delta\tau=0.01$ ns for the reduced dictionary in the delay domain. The channel tracking algorithm produces $N_{\rm est}=5$ estimated paths for each channel, and \textit{PathNet} \cite{chen2022joint,Yun_arXiv_2023} is used to determine the path orders. The paths classified as LOS or first order reflections (which matter for localization) are matched against their nearest true paths in the channels. The matching results are shown in Fig. \ref{Perfrom_ChanTrack} as an evaluation of the channel tracking performance. The errors of the estimated DoD and DoA can be limited to a maximum of $3\degree$ and $8\degree$. The accuracy is higher for DoD estimations because a larger array is employed at the RSU. The TDoA estimation errors are limited to $\leq 7$ ns. 
	
	The estimated channels combined with the initial location estimates are used for both training and testing V-ChATNet, so that the network can capture the noise features when using the proposed methods. The input sequence has a length of $C=16$, and the sample interval is set to $\cI=25$ to capture the channel temporal variations. V-ChATNet is trained with the batch size set to $64$, the learning rate set to $1\times 10^{-4}$, and the number of training epochs set to $1000$ with the early stopping strategy to avoid overfitting. Fig. \ref{example_traj} shows an example of the tracking results of a trajectory from the testing set, where the initial location estimate using the channel geometric transformation is denoted as ``Initial Location", and a \ac{KF} \cite{khalkhali2020vehicle} is also simulated for comparison. Results show that the  positions corrected by V-ChATNet are the most accurately aligned with the true trajectory. The \ac{CDF} of the localization accuracy can be observed in Fig. \ref{CDF_loc}. The initial localization based on the geometric system of equations yields the $5$, $50$, $80$, and $95$-th percentile 2D errors of $0.058$, $0.418$, $0.833$, and $1.611$ m. These values  decrease to $0.027$, $0.171$, $0.575$, and $1.574$ m when using the \ac{KF}, while V-ChATNet offers a significant improvement in the accuracy, reducing the errors up to $1.4$ m, with the percentile values of $0.014$, $0.065$, $0.120$, and $0.197$ m. Moreover, the network exhibits consistent performance on both the training and testing sets, indicating that it possesses reliable generalization capabilities.
	
	\section{Conclusion}
	We developed a system for joint  location and channel tracking  in realistic urban environments. The low complexity MOMP algorithm with reduced dictionaries enabled accurate channel tracking results, leading to an initial position estimate based on the solution of a system of geometric equations with an error below $\sim 1.61$ m  95\% of the time. Then, we developed V-ChATNet, an attention network  that takes the sequences of the historical channel estimates and the initial location estimates, and  extracts both the temporal and spatial features, providing a correction to the current location estimate. The experimental results show that our method achieves an accuracy of $0.2$ m for $95\%$ of the time for a vehicle driving at $60$ km/h in a realistic environment simulated by ray tracing.
	
	\bibliographystyle{IEEEtran}
	{
		\scriptsize
		\bibliography{refs}}

\begin{thebibliography}{10}
\providecommand{\url}[1]{#1}
\csname url@samestyle\endcsname
\providecommand{\newblock}{\relax}
\providecommand{\bibinfo}[2]{#2}
\providecommand{\BIBentrySTDinterwordspacing}{\spaceskip=0pt\relax}
\providecommand{\BIBentryALTinterwordstretchfactor}{4}
\providecommand{\BIBentryALTinterwordspacing}{\spaceskip=\fontdimen2\font plus
\BIBentryALTinterwordstretchfactor\fontdimen3\font minus
  \fontdimen4\font\relax}
\providecommand{\BIBforeignlanguage}[2]{{%
\expandafter\ifx\csname l@#1\endcsname\relax
\typeout{** WARNING: IEEEtran.bst: No hyphenation pattern has been}%
\typeout{** loaded for the language `#1'. Using the pattern for}%
\typeout{** the default language instead.}%
\else
\language=\csname l@#1\endcsname
\fi
#2}}
\providecommand{\BIBdecl}{\relax}
\BIBdecl

\bibitem{storck2020survey}
C.~R. Storck and F.~Duarte-Figueiredo, ``{A survey of 5G technology evolution,
  standards, and infrastructure associated with vehicle-to-everything
  communications by internet of vehicles},'' \emph{IEEE access}, vol.~8, pp.
  117\,593--117\,614, 2020.

\bibitem{wu2022compressive}
S.-H. Wu and G.-Y. Lu, ``Compressive beam and channel tracking with
  reconfigurable hybrid beamforming in {mmWave} {MIMO OFDM} systems,''
  \emph{IEEE Transactions on Wireless Communications}, 2022.

\bibitem{boljanovic2018tracking}
V.~Boljanovic, H.~Yan, and D.~Cabric, ``Tracking sparse {mmWave} channel under
  time varying multipath scatterers,'' in \emph{2018 52nd Asilomar Conference
  on Signals, Systems, and Computers}.\hskip 1em plus 0.5em minus 0.4em\relax
  IEEE, 2018, pp. 1274--1279.

\bibitem{zhu2020ray}
L.~Zhu, D.~He, B.~Ai, K.~Guan, S.~Dang, J.~Kim, H.~Chung, and Z.~Zhong, ``A ray
  tracing and joint spectrum based clustering and tracking algorithm for
  internet of intelligent vehicles,'' \emph{Journal of Communications and
  Information Networks}, vol.~5, no.~3, pp. 265--281, 2020.

\bibitem{chen2021millidegree}
Y.~Chen, L.~Yan, C.~Han, and M.~Tao, ``Millidegree-level direction-of-arrival
  estimation and tracking for terahertz ultra-massive {MIMO} systems,''
  \emph{IEEE Transactions on Wireless Communications}, vol.~21, no.~2, pp.
  869--883, 2021.

\bibitem{zhu2018high}
D.~Zhu, J.~Choi, Q.~Cheng, W.~Xiao, and R.~W. Heath, ``High-resolution angle
  tracking for mobile wideband millimeter-wave systems with antenna array
  calibration,'' \emph{IEEE Transactions on Wireless Communications}, vol.~17,
  no.~11, pp. 7173--7189, 2018.

\bibitem{zhao2018time}
J.~Zhao, H.~Xie, F.~Gao, W.~Jia, S.~Jin, and H.~Lin, ``Time varying channel
  tracking with spatial and temporal {BEM} for massive {MIMO} systems,''
  \emph{IEEE Transactions on Wireless Communications}, vol.~17, no.~8, pp.
  5653--5666, 2018.

\bibitem{burghal2019machine}
D.~Burghal, N.~A. Abbasi, and A.~F. Molisch, ``A machine learning solution for
  beam tracking in mmwave systems,'' in \emph{2019 53rd Asilomar Conference on
  Signals, Systems, and Computers}.\hskip 1em plus 0.5em minus 0.4em\relax
  IEEE, 2019, pp. 173--177.

\bibitem{koivisto2021channel}
M.~Koivisto, J.~Talvitie, E.~Rastorgueva-Foi, Y.~Lu, and M.~Valkama, ``Channel
  parameter estimation and {TX} positioning with multi-beam fusion in {5G}
  {mmWave} networks,'' \emph{IEEE Transactions on Wireless Communications},
  vol.~21, no.~5, pp. 3192--3207, 2021.

\bibitem{chu2022joint}
X.~Chu, Z.~Lu, D.~Gesbert, L.~Wang, X.~Wen, M.~Wu, and M.~Li, ``Joint vehicular
  localization and reflective mapping based on team {channel-SLAM},''
  \emph{IEEE Transactions on Wireless Communications}, vol.~21, no.~10, pp.
  7957--7974, 2022.

\bibitem{kim2022pmbm}
H.~Kim, K.~Granstr{\"o}m, L.~Svensson, S.~Kim, and H.~Wymeersch, ``{PMBM}-based
  {SLAM} filters in {5G} {mmWave} vehicular networks,'' \emph{IEEE Transactions
  on Vehicular Technology}, vol.~71, no.~8, pp. 8646--8661, 2022.

\bibitem{zeng2021massive}
X.~Zeng, F.~Zhang, B.~Wang, and K.~R. Liu, ``Massive {MIMO} for high-accuracy
  target localization and tracking,'' \emph{IEEE Internet of Things Journal},
  vol.~8, no.~12, pp. 10\,131--10\,145, 2021.

\bibitem{ge2022mmwave}
Y.~Ge, O.~Kaltiokallio, H.~Kim, J.~Talvitie, S.~Kim, L.~Svensson, M.~Valkama,
  and H.~Wymeersch, ``Mmwave mapping and {SLAM} for {5G} and beyond,''
  \emph{arXiv preprint arXiv:2211.16024}, 2022.

\bibitem{chen2022joint}
Y.~Chen, J.~Palacios, N.~Gonz{\'a}lez-Prelcic, T.~Shimizu, and H.~Lu, ``Joint
  initial access and localization in millimeter wave vehicular networks: a
  hybrid model/data driven approach,'' in \emph{2022 IEEE 12th Sensor Array and
  Multichannel Signal Processing Workshop (SAM)}.\hskip 1em plus 0.5em minus
  0.4em\relax IEEE, 2022, pp. 355--359.

\bibitem{vaswani2017attention}
A.~Vaswani, N.~Shazeer, N.~Parmar, J.~Uszkoreit, L.~Jones, A.~N. Gomez,
  {\L}.~Kaiser, and I.~Polosukhin, ``Attention is all you need,''
  \emph{Advances in neural information processing systems}, vol.~30, 2017.

\end{thebibliography}


\begin{thebibliography}{10}
\providecommand{\url}[1]{#1}
\csname url@samestyle\endcsname
\providecommand{\newblock}{\relax}
\providecommand{\bibinfo}[2]{#2}
\providecommand{\BIBentrySTDinterwordspacing}{\spaceskip=0pt\relax}
\providecommand{\BIBentryALTinterwordstretchfactor}{4}
\providecommand{\BIBentryALTinterwordspacing}{\spaceskip=\fontdimen2\font plus
\BIBentryALTinterwordstretchfactor\fontdimen3\font minus
  \fontdimen4\font\relax}
\providecommand{\BIBforeignlanguage}[2]{{%
\expandafter\ifx\csname l@#1\endcsname\relax
\typeout{** WARNING: IEEEtran.bst: No hyphenation pattern has been}%
\typeout{** loaded for the language `#1'. Using the pattern for}%
\typeout{** the default language instead.}%
\else
\language=\csname l@#1\endcsname
\fi
#2}}
\providecommand{\BIBdecl}{\relax}
\BIBdecl

\bibitem{Nokia5GNRPosWhitePaper2021}
R.~Keating, A.~Ghosh, B.~Velgaard, D.~Michalopoulos, and M.~S\"{a}ily, ``{The
  evolution of 5G New Radio positioning technologies},'' Nokia Bell Labs, Tech.
  Rep., 02 2021.

\bibitem{ShahmansooriTWC2018}
A.~Shahmansoori, G.~E. Garcia, G.~Destino, G.~Seco-Granados, and H.~Wymeersch,
  ``Position and orientation estimation through millimeter-wave {MIMO} in {5G}
  systems,'' \emph{IEEE Trans. Wireless Commun.}, vol.~17, no.~3, pp.
  1822--1835, 2018.

\bibitem{chen2022joint}
Y.~Chen, J.~Palacios, N.~Gonz{\'a}lez-Prelcic, T.~Shimizu, and H.~Lu, ``Joint
  initial access and localization in millimeter wave vehicular networks: a
  hybrid model/data driven approach,'' in \emph{IEEE12th Sensor Array and
  Multichannel Signal Processing Workshop (SAM)}, 2022, pp. 355--359.

\bibitem{Yun_arXiv_2023}
Y.~Chen, N.~González-Prelcic, T.~Shimizu, and H.~Lu, ``Learning to localize
  with attention: from sparse {mmWave} channel estimates from a single {BS} to
  high accuracy {3D} positioning,'' \emph{arXiv preprint}, 2023.

\bibitem{wu2022compressive}
S.-H. Wu and G.-Y. Lu, ``Compressive beam and channel tracking with
  reconfigurable hybrid beamforming in {mmWave} {MIMO OFDM} systems,''
  \emph{IEEE Transactions on Wireless Communications}, 2022.

\bibitem{zhu2020ray}
L.~Zhu, D.~He, B.~Ai, K.~Guan, S.~Dang, J.~Kim, H.~Chung, and Z.~Zhong, ``A ray
  tracing and joint spectrum based clustering and tracking algorithm for
  internet of intelligent vehicles,'' \emph{Journal of Communications and
  Information Networks}, vol.~5, no.~3, pp. 265--281, 2020.

\bibitem{chen2021millidegree}
Y.~Chen, L.~Yan, C.~Han, and M.~Tao, ``Millidegree-level direction-of-arrival
  estimation and tracking for {terahertz} ultra-massive {MIMO} systems,''
  \emph{IEEE Transactions on Wireless Communications}, vol.~21, no.~2, pp.
  869--883, 2021.

\bibitem{koivisto2021channel}
M.~Koivisto, J.~Talvitie, E.~Rastorgueva-Foi, Y.~Lu, and M.~Valkama, ``Channel
  parameter estimation and {TX} positioning with multi-beam fusion in {5G}
  {mmWave} networks,'' \emph{IEEE Transactions on Wireless Communications},
  vol.~21, no.~5, pp. 3192--3207, 2021.

\bibitem{chu2022joint}
X.~Chu, Z.~Lu, D.~Gesbert, L.~Wang, X.~Wen, M.~Wu, and M.~Li, ``Joint vehicular
  localization and reflective mapping based on team {channel-SLAM},''
  \emph{IEEE Transactions on Wireless Communications}, vol.~21, no.~10, pp.
  7957--7974, 2022.

\bibitem{kim2022pmbm}
H.~Kim, K.~Granstr{\"o}m, L.~Svensson, S.~Kim, and H.~Wymeersch, ``{PMBM}-based
  {SLAM} filters in {5G} {mmWave} vehicular networks,'' \emph{IEEE Transactions
  on Vehicular Technology}, vol.~71, no.~8, pp. 8646--8661, 2022.

\bibitem{palacios2022multidimensional}
J.~Palacios, N.~Gonz{\'a}lez-Prelcic, and C.~Rusu, ``Multidimensional
  orthogonal matching pursuit: theory and application to high accuracy joint
  localization and communication at mmwave,'' \emph{arXiv preprint
  arXiv:2208.11600}, 2022.

\bibitem{palacios2022low}
------, ``Low complexity joint position and channel estimation at millimeter
  wave based on multidimensional orthogonal matching pursuit,'' in \emph{2022
  30th European Signal Processing Conference (EUSIPCO)}.\hskip 1em plus 0.5em
  minus 0.4em\relax IEEE, 2022, pp. 1002--1006.

\bibitem{chaudhari2021attentive}
S.~Chaudhari, V.~Mithal, G.~Polatkan, and R.~Ramanath, ``An attentive survey of
  attention models,'' \emph{ACM Transactions on Intelligent Systems and
  Technology (TIST)}, vol.~12, no.~5, pp. 1--32, 2021.

\bibitem{khalkhali2020vehicle}
M.~B. Khalkhali, A.~Vahedian, and H.~S. Yazdi, ``Vehicle tracking with kalman
  filter using online situation assessment,'' \emph{Robotics and Autonomous
  Systems}, vol. 131, p. 103596, 2020.

\end{thebibliography}
	
\end{document}